\documentclass[aps,prl,floatfix,twocolumn]{revtex4}

\usepackage{amsmath}
\usepackage{units}
\usepackage{xspace}
\usepackage{graphicx}

\graphicspath{{figures/}}

\renewcommand{\vec}[1]{\boldsymbol{#1}}
\newcommand{\uvec}[1]{\hat{\boldsymbol{#1}}}
\newcommand{\avg}[1]{\left< #1 \right>}

\newcommand{\micron}{\unit{\mu m}\xspace}

\begin{document}

\title{Sun and Grier Reply (cond-mat.soft 0804.4632v1)}
\author{Bo Sun}
\author{David G. Grier}
\affiliation{Department of Physics and Center for Soft Matter Research,
New York University, New York, NY 10003}

\maketitle

Recently, Huang, Wu and Florin posted a Comment \cite{huang08} on our
preprint \cite{roichman08preprint} 
describing nonequilibrium circulation of a colloidal 
sphere trapped in a optical tweezer.
The Comment suggests that
evidence for toroidal probability currents 
obtained from experiments and simulations
in \cite{roichman08preprint} should be considered inconclusive.
The authors' concerns are based on two claims:
(1) that Brownian dynamics simulations of the trapped particle's
motions reveal no statistically significant circulation, and (2)
that a realistic description of the radiation
pressure acting on the trapped sphere is inconsistent with
the motion described in Ref.~\cite{roichman08preprint}.
In this Reply, we demonstrate both of these claims to be
incorrect, and thus the original results and conclusions in
Ref.~\cite{roichman08preprint} to be still valid.

\begin{figure}
  \centering
  \includegraphics[width=\columnwidth]{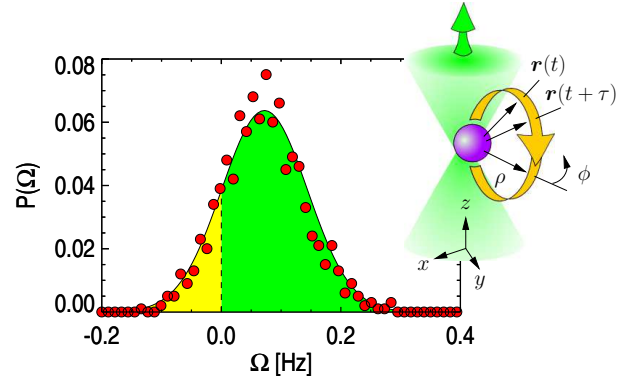}
  \caption{The probability $P(\Omega)$ compiled from
    1000 Brownian dynamics simulations for a particle
    to circulate in an optical tweezer at an average
    rate $\Omega$ over the course of 1000~\unit{s}.
    Inset:
    the experimental geometry, with a colloidal sphere
    localized near the focus of a beam of light propagating
    in the $\uvec{z}$ direction.  The broad arrow
    indicates circulation in the positive direction.
  }
  \label{fig:simulation}
\end{figure}

The system, shown schematically in Fig.~\ref{fig:simulation}, 
consists of a single colloidal sphere trapped in
a conventional optical tweezer formed by bringing
a beam of light to a diffraction-limited focus \cite{ashkin86}.
In Ref.~\cite{roichman08preprint}, we modeled the trap as
a radially symmetric harmonic well within which radiation pressure
exerts an additional force
directed along $\uvec{z}$:
\begin{equation}
  \label{eq:trap}
  \vec{F}_0(\vec{r}) 
  = 
  - k \, \vec{r}
  + f_1 \, \exp\left(- \frac{r^2}{2 \sigma^2} \right) \, \uvec{z}.
\end{equation}
The particle's position $\vec{r}$ is measured from the trap's focus,
$k$ is the trap's stiffness, $f_1$ sets the scale for the radiation
pressure, and $\sigma$ is
the effective range over which the focused light exerts forces on
the particle.
We assume that the particle is stably trapped, so that
$\epsilon = f_1 / (k\sigma)$ may be treated as a small parameter.

Were these the only forces acting on the sphere, the particle would
come to 
a stable mechanical equilibrium at a distance
$z_0 \approx \epsilon \sigma$ downstream of the focus.
The particle also is acted on by random thermal forces, however,
which displace it away from its equilibrium point.
Reference~\cite{roichman08preprint} demonstrates analytically
that the second term in Eq.~(\ref{eq:trap})
biases the trapped particle's thermal fluctuations in favor of
toroidal circulation in the sense depicted in Fig.~\ref{fig:simulation}.

Such a bias toward nonequilibrium circulation
would occur in any model for the radiation pressure
whose curl does not vanish.
The particular choice in Eq.~(\ref{eq:trap}) facilitates an
analytic treatment of the effect.
On this basis, we have claimed \cite{roichman08preprint}
that a particle trapped in an optical tweezer does come
to equilibrium, but rather acts as a Brownian motor
\cite{reimann96,reimann02,reimann02a}, with the nonconservative
component of $\vec{F}_0(\vec{r})$ biasing thermal fluctuations
in the manner of a thermal ratchet.

The authors of Ref.~\cite{huang08} do not call this result
into question, but rather claim that it is not conclusively
demonstrated by the simulations and experiments
presented in Ref.~\cite{roichman08preprint}.
To observe the predicted circulatory bias in a trajectory
$\vec{r}(t)$ discretely sampled over time intervals 
$\tau = 1/30~\unit{s}$, 
we introduced a measure of the mean circulation rate
\cite{roichman08preprint}
\begin{equation}
  \label{eq:omega}
  \Omega(t) = \frac{1}{2\pi} \,
  \frac{(\vec{r}(t+\tau) \times \vec{r}(t)) \cdot \uvec{\phi}}{
    \sqrt{
      \avg{
        (\rho - \avg{\rho})^2}
      \avg{
        (z - \avg{z})^2}
    }},\
\end{equation}
where $\vec{r} = (\rho,\phi,z)$ is measured in cylindrical
coordinates centered on the trap's focal point,
with $\uvec{z}$ pointing along the optical axis.
The predicted nonequilibrium circulation corresponds to clockwise
rotation in the $(\rho,z)$ plane, and to positive values of
$\Omega(t)$.

Huang \emph{et al.} claim \cite{huang08}
that Brownian dynamics simulations
corresponding to the experimental conditions in 
Ref.~\cite{roichman08preprint} show no statistically significant
trend in $\Omega(t)$, and thus no evidence for circulation.
Our numerical simulations, whose results are
presented in Fig.~\ref{fig:simulation},
demonstrate this claim to be incorrect.
Here, we have performed fourth-order Runge-Kutta integration of
a particle's trajectory evolving according to the Langevin
equation
\begin{equation}
  \label{eq:langevin}
  \gamma \dot{\vec{r}}(t) = \vec{F}_0(\vec{r}) + \vec{F}_1(t),
\end{equation}
where $\gamma = 6 \pi \eta a$ is the Stokes drag coefficient for
a sphere of radius $a$ moving through a fluid of viscosity $\eta$,
and $\vec{F}_1(t)$ is a zero-mean stochastic force whose variance
is the thermal energy scale.
Although the simulations in \cite{roichman08preprint} were performed
with the isotropic model in Eq.~(\ref{eq:trap}),
the authors of Ref.~\cite{huang08} generalize the harmonic restoring force
to account for
different trap stiffnesses in the three Cartesian directions,
using $k_x = 0.467~\unit{pN/\micron}$, $k_y = 0.4~\unit{pN/\micron}$
and $k_z = 0.08~\unit{pN/\micron}$.
In responding to their criticism, we adopt the same anisotropic
force law in simulating the motions of a sphere of radius
$a = 1.1~\micron$ in a trap of width $\sigma = a$.
Following Refs.~\cite{huang08} and \cite{roichman08preprint},
we also set $\epsilon = 0.1$ and adopted time steps of $10^{-4}~\unit{s}$.
Taking the suspending medium to be water at room temperature,
$\eta = 10^{-3}~\unit{Pa \, s}$.

Figure~\ref{fig:simulation} shows the distribution of mean
circulation rates, $\Omega = \avg{\Omega(t)}$, obtained
from 1000 independent runs, each of 1000~\unit{s} duration.
As Huang \emph{et al.} point out \cite{huang08}, individual
realizations can display either positive or negative circulation.
Contrary to their assertion, however, these variations do not 
occur with equal probability.
Out of 1000 realizations, only 131 showed negative circulation. 
A similarly small proportion of retrograde circulation
is observed experimentally.
The ensemble-averaged circulation rate 
$\Omega = 0.08~\unit{Hz}$
agrees quantitatively both with the experimental results and
also with the analytic predictions presented in
Ref.~\cite{roichman08preprint}.

Although the principal claim by Huang \emph{et al.} is thus
shown to be incorrect, their Comment raises the valid point
that the scattering force experienced by a colloidal sphere in
a real optical trap is likely to be more complicated than the
idealized model in Eq.~(\ref{eq:trap}), particularly for spheres
larger than the wavelength of light.
The detailed form of $\vec{F}_0$ is less important, however,
than the presence of a rotational component, 
$\nabla \times \vec{F}_0 \neq 0$, for biasing the system
out of equilibrium.
It is this rotational
component that breaks the spatiotemporal symmetry of the
particle's fluctuations to create a net flux in its probability
density \cite{reimann96}.
The particular form in Eq.~(\ref{eq:trap}) was selected
more for its analytic tractability 
than for its accuracy as a
model for radiation pressure in optical traps.

The form for the scattering force presented in Ref.~\cite{huang08}
also has a rotational component, and so will give rise to
circulation in the particle's trajectory.
Unlike the scattering force in Eq.~(\ref{eq:trap}), which is peaked
on the optical axis, the ray-optics calculation in Ref.~\cite{huang08}
increases with distance from the optical axis, and so would
induce retrograde circulation.
This observation raises the interesting point that
optically trapped particles' behavior may
be more complicated than is predicted by the idealized model
in Eq.~(\ref{eq:trap}).

If this model for the radiation pressure were relevant to the experiments 
in Ref.~\cite{roichman08preprint}, then the measured trajectories
also should have displayed retrograde circulation.
Huang \emph{et al.} suggest that the discrepancy can be ascribed
to insufficient statistics in the experimental analysis.
We argue instead that the result for the scattering force presented 
in Fig.~2 of Ref.~\cite{huang08} reflects only the $\uvec{z}$
component of the optical force that a particle would experience
at the optical tweezer's focal point.
In fact, the silica sphere is more than twice as dense as the water
in which it is suspended, and so settles roughly $2~\unit{\mu m}$
\emph{below} the focal point.
In this region of the beam, the total optical force
computed by a fully vectorial theory \cite{nieminen07}
has a uniformly positive curl.
Consequently, the particle should undergo positive circulation, as reported.
This is not to say that Eq.~(\ref{eq:trap}) is an accurate representation
for the optical forces experienced by the sphere, but rather that
the form proposed in Fig.~2 of Ref.~\cite{huang08} is not.

In conclusion, we have demonstrated that the concerns raised by
Huang, Wu and Florin in Ref.~\cite{huang08}
can be ascribed to inadequate statistical analysis of their simulations
and to an incomplete analysis of the scattering force
acting on optically trapped spheres.
The results and conclusions presented in
Ref.~\cite{roichman08preprint} therefore remain unchanged.

This work was supported by the National Science Foundation through
Grant Number DMR-0606415.  B.S.\ acknowledges support of a Kessler
Family Foundation Fellowship.


\begin{thebibliography}{7}
\expandafter\ifx\csname natexlab\endcsname\relax\def\natexlab#1{#1}\fi
\expandafter\ifx\csname bibnamefont\endcsname\relax
  \def\bibnamefont#1{#1}\fi
\expandafter\ifx\csname bibfnamefont\endcsname\relax
  \def\bibfnamefont#1{#1}\fi
\expandafter\ifx\csname citenamefont\endcsname\relax
  \def\citenamefont#1{#1}\fi
\expandafter\ifx\csname url\endcsname\relax
  \def\url#1{\texttt{#1}}\fi
\expandafter\ifx\csname urlprefix\endcsname\relax\def\urlprefix{URL }\fi
\providecommand{\bibinfo}[2]{#2}
\providecommand{\eprint}[2][]{\url{#2}}

\bibitem[{\citenamefont{Huang et~al.}()\citenamefont{Huang, Wu, and
  Florin}}]{huang08}
\bibinfo{author}{\bibfnamefont{R.}~\bibnamefont{Huang}},
  \bibinfo{author}{\bibfnamefont{P.}~\bibnamefont{Wu}}, \bibnamefont{and}
  \bibinfo{author}{\bibfnamefont{E.-L.} \bibnamefont{Florin}},
  \emph{\bibinfo{title}{Comment on "influence of non-conservative optical
  forces on the dynamics of optically trapped colloidal spheres: The fountain
  of probability", cond-mat.soft 0804.0730v1}}, \eprint{cond-mat.soft
  0804.4632v1}.

\bibitem[{\citenamefont{Roichman et~al.}()\citenamefont{Roichman, Sun,
  Stolarski, and Grier}}]{roichman08preprint}
\bibinfo{author}{\bibfnamefont{Y.}~\bibnamefont{Roichman}},
  \bibinfo{author}{\bibfnamefont{B.}~\bibnamefont{Sun}},
  \bibinfo{author}{\bibfnamefont{A.}~\bibnamefont{Stolarski}},
  \bibnamefont{and} \bibinfo{author}{\bibfnamefont{D.~G.} \bibnamefont{Grier}},
  \emph{\bibinfo{title}{Influence of non-conservative optical forces on the
  dynamics of optically trapped colloidal spheres: The fountain of
  probability}}, \eprint{cond-mat.soft 0804.0730v1}.

\bibitem[{\citenamefont{Ashkin et~al.}(1986)\citenamefont{Ashkin, Dziedzic,
  Bjorkholm, and Chu}}]{ashkin86}
\bibinfo{author}{\bibfnamefont{A.}~\bibnamefont{Ashkin}},
  \bibinfo{author}{\bibfnamefont{J.~M.} \bibnamefont{Dziedzic}},
  \bibinfo{author}{\bibfnamefont{J.~E.} \bibnamefont{Bjorkholm}},
  \bibnamefont{and} \bibinfo{author}{\bibfnamefont{S.}~\bibnamefont{Chu}},
  \bibinfo{journal}{Opt. Lett.} \textbf{\bibinfo{volume}{11}},
  \bibinfo{pages}{288} (\bibinfo{year}{1986}).

\bibitem[{\citenamefont{Reimann}(2002)}]{reimann02}
\bibinfo{author}{\bibfnamefont{P.}~\bibnamefont{Reimann}},
  \bibinfo{journal}{Phys. Rep.} \textbf{\bibinfo{volume}{361}},
  \bibinfo{pages}{57} (\bibinfo{year}{2002}).

\bibitem[{\citenamefont{Reimann et~al.}(2002)\citenamefont{Reimann, Van~den
  Broeck, Linke, {H\"anggi}, Rubi, and {P\'erez-Madrid}}}]{reimann02a}
\bibinfo{author}{\bibfnamefont{P.}~\bibnamefont{Reimann}},
  \bibinfo{author}{\bibfnamefont{C.}~\bibnamefont{Van~den Broeck}},
  \bibinfo{author}{\bibfnamefont{H.}~\bibnamefont{Linke}},
  \bibinfo{author}{\bibfnamefont{P.}~\bibnamefont{{H\"anggi}}},
  \bibinfo{author}{\bibfnamefont{J.~M.} \bibnamefont{Rubi}}, \bibnamefont{and}
  \bibinfo{author}{\bibfnamefont{A.}~\bibnamefont{{P\'erez-Madrid}}},
  \bibinfo{journal}{Phys. Rev. E} \textbf{\bibinfo{volume}{65}},
  \bibinfo{pages}{031104} (\bibinfo{year}{2002}).

\bibitem[{\citenamefont{Reimann et~al.}(1996)\citenamefont{Reimann, Bartussek,
  Haussler, and H\"{a}nggi}}]{reimann96}
\bibinfo{author}{\bibfnamefont{P.}~\bibnamefont{Reimann}},
  \bibinfo{author}{\bibfnamefont{R.}~\bibnamefont{Bartussek}},
  \bibinfo{author}{\bibfnamefont{R.}~\bibnamefont{Haussler}}, \bibnamefont{and}
  \bibinfo{author}{\bibfnamefont{P.}~\bibnamefont{H\"{a}nggi}},
  \bibinfo{journal}{Phys. Lett. A} \textbf{\bibinfo{volume}{215}},
  \bibinfo{pages}{26} (\bibinfo{year}{1996}).

\bibitem[{\citenamefont{Nieminen et~al.}(2007)\citenamefont{Nieminen, Y.,
  Stilgoe, Knoner, Branczyk, Heckenberg, and Rubinsztein-Dunlop}}]{nieminen07}
\bibinfo{author}{\bibfnamefont{T.~A.} \bibnamefont{Nieminen}},
  \bibinfo{author}{\bibfnamefont{L.~V.~L.} \bibnamefont{Y.}},
  \bibinfo{author}{\bibfnamefont{A.~B.} \bibnamefont{Stilgoe}},
  \bibinfo{author}{\bibfnamefont{G.}~\bibnamefont{Knoner}},
  \bibinfo{author}{\bibfnamefont{A.~M.} \bibnamefont{Branczyk}},
  \bibinfo{author}{\bibfnamefont{N.~R.} \bibnamefont{Heckenberg}},
  \bibnamefont{and}
  \bibinfo{author}{\bibfnamefont{H.}~\bibnamefont{Rubinsztein-Dunlop}},
  \bibinfo{journal}{J. Opt. A} \textbf{\bibinfo{volume}{9}},
  \bibinfo{pages}{S196} (\bibinfo{year}{2007}).

\end{thebibliography}

\end{document}